\newcommand{\rem}[1]{}

\documentclass[a4paper,12pt]{article}
\usepackage{epsfig}
\usepackage{hyperref}

\begin{document}


\title{ Low-energy unphysical saddle  in polynomial molecular potentials}
\author{Alessio Del Monte$^1$, Nicola Manini$^{1,2}$,
\\
        Luca Molinari$^{1,3}$\footnote{Author for correspondence. e-mail:
	  luca.molinari@mi.infn.it}\ \ and Gian Paolo Brivio$^{4}$
\\ \small 
$^1$Dipartimento di Fisica,
Universit\`a degli Studi di Milano,
\\ \small 
Via Celoria 16, I-20133 Milano, Italy
\\ \small 
$^2$INFM, Unit\`a di Milano, Via Celoria 16, Milano, Italy
\\ \small 
$^3$INFN, Sezione di Milano, Via Celoria 16, Milano, Italy
\\ \small 
$^4$Dipartimento di Scienza dei Materiali,
\\ \small 
Universit\`a di Milano-Bicocca, Via Cozzi 53, I-20125 Milano,
\\ \small 
and INFM, Unit\`a di Milano-Bicocca, Milano, Italy
}
\date{Oct.29, 2004}
\maketitle
\begin{abstract}
Vibrational spectra of polyatomic molecules are often obtained from a
polynomial expansion of the adiabatic potential around a minimum.  For
several molecules, we show that such an approximation displays an
unphysical saddle point of comparatively small energy, leading to a region
where the potential is negative and unbounded. This poses an upper limit
for a reliable evaluation of vibrational levels. We argue that the presence
of such saddle points is general.
\end{abstract}

\section{\label{sec:level1}Introduction}
%
The morphology of the adiabatic potential energy surface (APES), especially
its low-energy minima and saddle points, is at the basis of the quantum
chemistry of reaction paths and conformational transitions \cite{Ramquet00}.
The adiabatic potential governs the low-energy vibrational dynamics of a
rigid molecule of $N$ atoms and is a complicate function of $d=3N\!-\!6$
internal coordinates, such as bond lengths, bending or torsion angles
($d=3N\!-\!5$ for linear molecules).
Its actual determination is usually a very demanding
problem \cite{Herman99}, as the information contents of a function of $d$
coordinates grows exponentially with $d$.
{\em Global} parametrizations of the APES computed with ab-initio methods
are presently accessible for small molecules
\cite{Kuhn99,Leforestier01,Carter02,Kurkal03}.
A standard {\em local} parametrization, which was the mainroad to
vibrational dynamics and is now mostly employed for medium-sized
molecules\cite{Maslen92}, is provided by a Taylor expansion around the
global minimum of the APES.
For a deep minimum, one can isolate the vibrational contribution to the
Hamiltonian, and expand it in normal modes\cite{Wilson55}:
\begin{eqnarray} 
H &=& T+V = \frac{1}{2}\sum_a  \hbar \omega_a (p_a^2 + q_a^2 )
\nonumber \\\label{Taylor}
&&
+ \frac{1}{3!}\sum_{abc} \phi_{abc}\,q_a q_b q_c 
+ \frac{1}{4!}\sum_{abcd} \phi_{abcd}\,q_a q_b q_c q_d +\ldots
\end{eqnarray}
$q_a$ are the dimensionless normal-mode coordinates measured with respect to 
the equilibrium geometry, and $p_a$ are the conjugate momenta.
The frequencies $\omega_a$ and higher-order constants $\phi_{ab...}$ of
several molecules have been computed {\it ab-initio}, and sometimes refined
by comparison with spectroscopic data \cite{Lee97}.
For a simple molecule such as water, the anharmonic constants have been
determined up to sixth order \cite{Csaszar97}, and for several polyatomic
molecules the literature reports calculations of third and fourth-order
constants
\cite{Cizek93,Martin95,Koput98,Miani00benz,Burcl03,Demaison03,Yagi04}.
It is widely recognized that the Taylor series should be used with caution,
since it only converges within a limited radius \cite{Steele62}.
The truncated series provides a locally accurate parametrization of the
actual APES.
Hereafter, we shall refer to the (finite) polynomial potential $V(q)$ in
equation~(\ref{Taylor}) as to the PP.

We consider the available data for the PP of several molecules. In all
cases we find a saddle point of comparatively small energy, leading to an
unphysical dissociative region where the potential is not bounded from
below.
We also find that, for most molecules, the potential well accomodates very
few quantum levels up to the saddle energy.
To our surprise, this problem seems to be neglected: its analysis is the
main purpose of the present study.

A main use of the polynomial expansion (\ref{Taylor}) is the
calculation of a number of vibrational levels, usually by means of
perturbation or variational theories, or by numerical diagonalization.
Contrary to finite-order perturbative calculations, a non-perturbative
determination of vibrational levels inevitably detects tunneling to the
unphysical region through the saddle.
We show that sharp vibrational levels do occur only in the energy
region below a quantum energy border, given by the sum of the energy of the
lowest saddle (the classical border) plus the zero-point energy of the
$d-1$ ``transverse'' modes (of positive curvature) at the saddle point.

The paper is organized as follows: In Sec.~\ref{classical} we present the
data for the lowest unphysical saddle point of several molecules, based on
{\it ab-initio} anharmonic constants available in the literature, and give 
arguments for the
general occurrence of a low-energy saddle in polynomial approximations of
the molecular APES. 
In Sec.~\ref{quantum} we introduce the quantum energy border for the 
tunneling regime, and illustrate its influence on the vibrational spectra 
of two molecules which respond very differently, water and ketene. 
We employ a non-perturbative method developed by us for evaluating the 
spectra, which is presented in the Appendix.  In 
Sec.~\ref{discussion} we discuss the results and the ensuing scenario.

\section{The unphysical saddle}
\label{classical}

A unique feature of the polynomial expansion in equation~(\ref{Taylor}) for
the special case of diatomic molecules ($d\!=\!1$) is that many even-power
terms are {\em positive} (for example all even-power terms are
positive for the Morse and Lennard-Jones functions).
As a consequence, the truncation of the series to a positive even-power
coefficient gives a lower-bounded potential, and thus a well-defined
quantum problem, characterized by an infinite set of discrete levels.
This feature of $d\!=\!1$ is unfortunately lost when the power expansion
(\ref{Taylor}) is extended to polyatomic molecules ($d>1$): in all systems
which we could obtain the anharmonic parameters for, we verified the
occurrence of regions where the fourth-order PP is unbounded below.

\centerline{ \bf [Insert figure 1 about here]}

Energy barriers separate different minima of a {\em physical} APES, 
corresponding to different local equilibrium configurations (isomers) of 
the molecule.
The isomerization dynamics occurs mainly {\it via} quantum tunneling or
thermal activation through the lowest saddle of the barrier
\cite{Carrington84,Hanggi90}.
Likewise, for the PP, energy barriers separate the region of the physical
minimum around which the expansion is based and well grounded, from the
{\em unphysical} regions where the potential drops to $-\infty$.
The escape to the unphysical region is driven by the lowest saddle, which
introduces an energy ``border'' that limits the range for (meta-stable)
quantum levels allowed in the physical well of the PP.
Figure~\ref{oneD:fig} illustrates this concept in a simple
$d\!=\!1$-dimensional context, where we purposedly truncated the polynomial
expansion (\ref{Taylor}) to an odd order.
It is clear that the saddle lies in a region where the polynomial has
already become a poor approximation to the actual APES.

To determine the lowest saddle point of the PP of a polyatomic molecule, we
first locate all stationary points in the neighborhood of ${\bf q}\!=\!{\bf
0}$.
For this purpose, we compute analytically the gradient $\nabla_{\bf q}V$,
and repeatedly solve the polynomial equation $\nabla_{\bf q}V={\bf 0}$ by
the Newton method, starting from a scattered set of several thousand random
initial points. This procedure generates a large number of stationary points
of the PP.
We then select the solution ${\bf q}_{\rm s}$ with the lowest
positive energy $E_{\rm s}$ and check that it is indeed a saddle point,
characterized by one negative and $d-1$ positive curvatures.
Finally we verify that the PP, restricted to the straight
line through the points ${\bf q}\!=\!{\bf 0}$ and ${\bf q}_{\rm s}$, has a
shape qualitatively similar to the dashed line of figure~\ref{oneD:fig},
i.e.\ that tunneling indeed occurs through a single barrier to a region
where the potential drops to $-\infty$.

\centerline{ \bf [Insert table 1 about here]}

For several molecules, table 1\ 
reports the height $E_{\rm s}$ of the lowest (unphysical) saddle point of
the PP, measured from the bottom of the potential well.
Surprisingly, these saddles are low: only about few times a typical
harmonic vibrational frequency of the molecule. As a result, few, if
any, vibrational states sit below $E_{\rm s}$.
The reported values of $E_{\rm s}$ of the diatomic molecules
\cite{Bransden} set the
characteristic scale given by the bond anharmonicities, in the
$10^4$~cm$^{-1}$ region.
This can be regarded as an upper bound, as off-diagonal anharmonicities of
the PP can only provide lower saddles, involving several normal coordinates
at the same time.
Indeed table 1\ 
shows lower saddles for polyatomic molecules of increasing number of atoms.
Especially low saddles are found for molecules characterized by soft
torsional modes, such as methanol CH$_3$OH.
The difference in energy between the two saddle heights obtained from
different available PP of H$_2$O suggests an estimate of the accuracy of
the reported values of $E_{\rm s}$ due to the approximations involved in
{\it ab-initio} calculations of the anharmonic parameters.

The occurrence of a saddle leading to an unphysical region is by no means
specific of the PP of the molecules listed in table~1
: we
argue that this is a general feature to be found in the PP of most
polyatomic molecules.
Indeed, also for an even-power truncation, the PP can easily drop to
$-\infty$ in some direction in ${\bf q}$ space
\footnote{
For example, the fourth-order terms $ \phi_{1111} q_1^4 + \phi_{1222} q_1
q_2^3 + \phi_{2222} q_2^4 $ combine to $(\gamma^4 \phi_{1111}
+\gamma\phi_{1222} +\phi_{2222})\, q_2^4$ along the line $q_1=\gamma
\,q_2$, and the numeric coefficient $\gamma^4 \phi_{1111} + \gamma
\phi_{1222} +
\phi_{2222}$ can easily be negative, provided that $|\phi_{1222}|$ is large
enough and that $\gamma$ is chosen suitably.  Also, to make things worse,
even though fully diagonal $\phi_{aaaa}$ terms are usually positive, there
often occur semi-diagonal terms $\phi_{aabb}$ with negative sign.
}.
The precise value of the 4th-order parameters $\phi_{abcd}$ (including
their sign) is determined by the {\em local} properties of the physical APES at
its minimum ${\bf q}={\bf 0}$, not by any requirement of confining behaviour
at large distance: the PP of a real molecule easily contains negative
semi-diagonal terms $\phi_{aabb}$ and sizable mixed terms $\phi_{abcd}$,
which in turn produce regions where the PP drops to $-\infty$.
In practice, the same argument prevents confining behaviour also of 6th and
higher even-order terms, and an even-power truncated PP does not
behave any better, away from the physical minimum, than an odd-power
truncated PP.
Therefore, we consider it extremely unlikely (though technically possible)
that a real polyatomic molecular potential may ever be found whose
polynomial expansion at the minimum (truncated at any order $>2$) is
lower-bounded everywhere.
Several methods to circumvent the incorrect behavior of polynomial
approximations at large $\mathbf q$ are commonly employed: for example,
Morse coordinates for stretching modes are used to correct the unphysical
power-law asymptotics of the PP \cite{Murrell87,Carter93,Zhou02}.

\section{The quantum energy border}
\label{quantum}

As the PP has no lower bound, the associated Schr\"odinger problem is
ill-defined. However, resonant states with complex energy values
$E_a-i\Gamma_a$ usually exist in the well \cite{Miller80,Carrington84,Hanggi86,Benderskii94,Zamastil01,Miller03,Yaris78,Rescigno97}.
Deep in the well, $\Gamma_a$ is in general exponentially small and
proportional to the Gamow factor $e^{-2 S/\hbar }$, where $S$ is the
imaginary-time action along the most probable tunneling path inside the
barrier.
Resonances appear as sharp peaks in the spectral density, at energies
$E_a$ extremely close to the eigenenergies of the Schr\"odinger equation
restricted to the well.
As energy increases toward the threshold value, tunneling evolves into a
``leaking'' regime, characterized by the appearence of resonances whose
imaginary part $\Gamma_a$ is comparable to the real part \cite{Yaris78},
relics of further excited states in the well, strongly hybridized to the
continuum.

In one dimension, the threshold coincides with the saddle energy $E_{\rm s}$.
For $d>1$, an effect absent in the simple one-dimensional
picture of figure~\ref{oneD:fig} is to be considered: tunneling through the
barrier at the saddle point is hindered by the ``transverse'' motion of the
degrees of freedom perpendicular to the one crossing the barrier.
These perpendicular degrees of freedom are associated to a minimum energy
$E_{\rm zp}^\perp$ due to Heisenberg's uncertainty, that adds to the saddle
height to determine the quantum energy border between the tunneling and the
leaking regimes
\begin{equation}
\label{Eqb}
E_{\rm qb} = E_{\rm s}+E_{\rm zp}^\perp
\,.
\end{equation}
As the study of tunneling problems \cite{Kitamura00,Miani00} suggests, we
approximate $E_{\rm zp}^\perp$ by its harmonic expression
\begin{equation}
\label{Ezp}
E_{\rm zp}^\perp\simeq \frac 12 \sum_{i=2}^d \hbar \omega'_i
\end{equation}
in terms of the $d-1$ real harmonic frequencies $\omega'_i$ at the saddle
point ($\omega'_1$ is the imaginary frequency associated to the tunneling
direction).
Since, for most polyatomic molecules, $E_{\rm zp}^\perp$ is close to the 
zero-point energy of the ground state $E_{\rm zp}({\bf 0})$ (see
table~1
), the raising of the energy border due to 
$E_{\rm zp}^\perp$ recovers a spectral range ($E_{\rm zp}({\bf 0}) < E 
< E_{\rm qb}$)
where quasistationary vibrational levels are to be found, of extension
comparable to the classical region for bounded motion ($0\leq E \leq E_{\rm
s}$).

To illustrate the difficulties brought in by the unphysical saddle in
polynomial approximations of APES, we compute non-perturbatively the
vibrational spectra of two molecules, water and ketene, with different
outcomes.
For both, we employ {\it ab-initio} quartic anharmonic potentials for the
normal modes \cite{Csaszar97,East95}: while the PP of the water molecule
features an energy range between the zero point energy and the quantum
border containing a number of vibrational levels, for ketene the range is
so small to contain very few levels.
%
We evaluate the levels $|E_a\rangle $ of the vibrational Hamiltonian $H$
(\ref{Taylor}) by means of a non-perturbative technique equivalent to an
exact diagonalization on a finite basis.  Its advantage over standard
Lanczos/Davidson diagonalization is a uniform precision throughout the
whole spectrum.
The Green function 
\begin{equation}\label{green:eq}
G(E+i\varepsilon)=\langle {\bf v}_0|
(E+i\varepsilon -H)^{-1}|{\bf v}_0\rangle
\end{equation}
is computed with an iterative procedure on a set of values $E$. The initial
state $|{\bf v}_0\rangle $ is chosen as an eigenstate of the harmonic part
of $H$.  The spectrum is obtained from the lineshape function
\begin{eqnarray}
I(E)&=& -{1\over \pi}{\rm Im}\,G(E+i\varepsilon) 
=\frac{\varepsilon}{\pi} \sum_a 
 \frac{|\langle E_a|{\bf v}_0\rangle |^2}{(E-E_a)^2+\varepsilon^2 }
\,.
\end{eqnarray}
Eigenvalues show up as peaks of $I(E)$, with heights proportional to the
squared overlap of the exact eigenstates to the initial excitation.  The
parameter $\varepsilon$ introduces a phenomenological Lorentzian
broadening of the lines $E_a$.

A few words on the method for evaluating the Green function are necessary.
First, one constructs a basis of harmonic states grouped into families $T_i$
(tiers), $i=0\ldots N$, adapted to the specific PP and 
the initial 
state $|{\bf v}_0\rangle$ 
\cite{Bixon68,Uzer91,Marshall91}.
Next, the Green function is evaluated through the exact
recursive relation equation~(\ref{genericinv}), as detailed in the Appendix.
As the basis is finite, the spectrum consists
of a finite number of real eigenvalues.
By evaluating the spectra $I(E)$ with incresing number of tiers $N$, one has
a control of convergence in the regions below the quantum border, where
tunneling is suppressed. Above the border, no convergence is expected as
the basis size is changed.

\subsection{Water} 
\label{h2o}

We employ the {\it ab-initio} quartic PP parameters listed in table 6 of
Ref.~\cite{Csaszar97}.
%
The two real frequencies $\hbar\omega'_2=4758$\,cm$^{-1}$ and
$\hbar\omega'_3=4849$\,cm$^{-1}$ at the lowest saddle point of H$_2$O
($E_{\rm s}=6846$\,cm$^{-1}$) are found so much larger than the three
curvatures at the minimum, that $E_{\rm zp}^\perp> E_{\rm zp}({\bf 0})$
(table~1
): this produces a rather narrow saddle, which
pushes the quantum border up to $E_{\rm qb}=11649$\,cm$^{-1}$.

\centerline{ \bf [Insert figure 2 about here]}

We take the harmonic fundamental excitation of the $\omega_1$ (symmetric OH
stretching) mode as initial state $|{\bf v}_0\rangle$, and obtain the
spectrum in figure~\ref{water_conv:fig}.
The peaks, representing exact vibrational levels of the PP, are assigned to
the harmonic quantum numbers of the closest state resulting from standard
second-order perturbation theory.
Convergence is studied by increasing the tier number from $N=3$ to 
$N=15$. It is very good already using $N=3$ tiers,
thus showing the effectiveness of the tiers Green-function method.
%
%
Hardly any tunneling is observed below $E_{\rm qb}$.
Across $E_{\rm qb}$, the appearence of the weak non-converging satellites
confirms that the resonances in this region are affected by detectable
leakage to the continuum.
Indeed, since the lowest saddle lies in a direction involving mainly mode
1, the chosen initial state $|1,0,0\rangle$ promotes tunneling.
%
Nonetheless, the stabililty of several features even above $E_{\rm qb}$
indicates that a number of fairly long-lived physical states are so localized
in the well that their overlap to the outside continuum is relatively small.
%

\subsection{Ketene}
\label{h2c2o}
As shown in table 1, according to the {\it ab-initio} calculation of
Ref.~\cite{East95} the lowest saddle of the PP of ketene H$_2$C$_2$O is
very low.
The resulting useful energy range is therefore extremely narrow, from
$E_{\rm zp}({\mathbf 0})\simeq 6900$~cm$^{-1}$ to $E_{\rm qb} \simeq
7500$~cm$^{-1}$.
Only the ground state and the fundamental excitations of the lowest bending
modes are located in the spectral region below the quantum border.
%
The PP of ketene represents therefore a particulary unfavorable case where
no convergence is expected for any excitation.

\centerline{ \bf [Insert figure 3 about here]}

Figure \ref{ketene_conv:fig} shows the spectral region from the ground
state to the CH stretching mode, obtained with the $a_1$ initial excitation
$|1,0,0,0,0,0,0,0,0\rangle$ of the CH symmetric stretching mode $\nu_1$,
for different sizes of the basis.
In this region, only a few fundamental and overtone/combination states of
$a_1$ symmetry should be found.
Instead, tens of spurious structures arise, which show no tendency to
converge as the basis size increases.
This is to be contrasted with the converging spectral range of H$_2$O,
where a stable spectrum is achieved already with a small basis of $N=3$
tiers.
%

\section{Discussion}
\label{discussion}

We show that polynomial approximations of molecular potentials usually
display unphysical saddles that lead to regions where the potential is not
lower-bounded.
The height of the saddle and the zero-point energy of transverse modes both
determine a quantum energy border.
Unless the saddle is very low, as e.g.\ in the PP of the ketene molecule,
in the spectral region below the quantum border tunneling is exponentially
small and standard perturbative treatment of the anharmonic interactions
usually provides a good approximation to the sharp level positions.
Perturbative calculations are often extended to higher energies, based on the
general belief that, like in the $d\!=\!1$ Morse case, the second-order
approximation compensates the wrong behaviour of the PP away from the
minimum and reproduces the levels of a physical APES \cite{Mills85}.

Our non-perturbative calculations show that, as energy is raised above 
the quantum border, resonances leave the
tunneling regime and couple more and more strongly to the continuum,
practically washing out all spectral details.
Such highly excited states are of scarce physical relevance anyway, since
the extension of the associated wavefunction explores regions where the PP
becomes a very poor approximation of the true APES.

The present results cast a shadow on the traditional use of the
polynomial approximation of the physical APES for the calculation of highly
excited vibrational or roto-vibrational spectra of polyatomic molecules.
For example, in most molecules, IVR (intramolecular vibrational relaxation)
spectra involve energy levels much above the saddle \cite{Pearman98}.
Hence, IVR applications of exact numerical methods, such as the Lanczos or
Davidson algorithms \cite{Gruebele96,Wyatt98,Pochert00,Callegari03} or the
Green-function method proposed here, are bound to face the problem of the
saddle of the PP. The consequent broadening of the resonant states poses an
intrinsic limit to the spectral accuracy which can be obtained.
A real progress may be achieved by the use of smarter local
parametrizations of the APES (e.g.\ based on Morse coordinates
\cite{Carter93,Zhou02}).

\section*{Acknowledgements}

We thank J.H. van der Waals, G. Scoles, K. Lehmann, A. Callegari for useful
discussions, and J.M.L. Martin, J.P. Fran\c{c}ois for kindly providing us
the PP parameters for ethylene.

\appendix
\section{Non-perturbative evaluation of the Green function.}

We present a general procedure, inspired to Ref.~\cite{Stuchebrukhov93} 
and there indicated as ``tiers method'', for the non-perturbative 
evaluation of the eigenvalues and spectral weights of a Hamiltonian 
decomposed as $H=H_0+H'$.
%
For definiteness, we consider the problem at hand: $H_0=\sum \hbar\omega_b
(a_b^\dagger a_b+1/2)$ describes $d$ independent oscillators, with
eigenvectors $|{\bf v}\rangle =|v_1,v_2,\ldots, v_d\rangle$, $H'$ is the
anharmonic part of the potential (\ref{Taylor}).

The first step of the method is to partition the unperturbed eigenvectors
in families (tiers) $T_0,\ T_1,\ldots $ of decreasing perturbative
relevance, such that the matrix representation of $H$
is block-tridiagonal in the tiers. Symbolically we write the blocks as
$H_{ii}=\langle T_i|H|T_i\rangle $ and $H_{i,i+1}=\langle T_i|H|T_{i+1}
\rangle $.
Depending on the problem under investigation, a set of unperturbed
states $|{\bf v}_{0,\alpha}\rangle $ ($\alpha =1,\ldots ,t_0)$ is selected to
form the initial tier $T_0$.
In the computations of this paper, $T_0$ contains a single initial state.
The action of $H'$ on $T_0$ gives new vectors: the basis states that have
non-zero overlap with them, and are not in $T_0$, are collected in $T_1$.
We label them as $|{\bf v}_{1,\alpha}\rangle $ ($\alpha =1\ldots t_1)$.
For a finite set $T_0$, and an interaction $H'$ which is a polynomial in the
raising and lowering operators, tier $T_1$ and subsequent ones are
finite.
Next we consider the set $H'T_1$, and expand it in the
eigenvectors already in $T_0$ and $T_1$, plus new ones that are collected
in tier $T_2$. The process is iterated to produce further tiers 
$T_3$, $T_4$, ...
Up to this point the method is simply a smart algorithm to generate
systematically a good approximate basis for a quantum problem where some
non-interacting part $H_0$ of the Hamiltonian can be singled out.
Indeed, similar basis generation has been employed successfully in
different contexts
\cite{Stuchebrukhov93,Wyatt98,Callegari03,Manini03,Wang03}.
However, the hierarchical basis structure and the corresponding 
block-tridiagonal form of the Hamiltonian, suggest a natural iterative
method to construct the spectrum.

The matrix element of the resolvent ${\cal G}(z)=(zI-H)^{-1}$ in $T_0$ is
precisely the Green function in equation~(\ref{green:eq}).
We propose to compute it based on the following formula for the inversion
of partitioned matrices, with square diagonal blocks (we omit unneeded
terms):
\begin{equation}
M= \left(
\matrix{
M_{11} & M_{12}\cr M_{21} & M_{22}
}
\right)\,, \quad
M^{-1}= \left(
\matrix{
[M_{11}-M_{12}(M_{22})^{-1}M_{21}]^{-1} & \ldots \cr \ldots & \ldots 
}  \,.
\right)
\label{inv}
\end{equation}
We apply this formula by identifying $M$ with the matrix representations 
of $zI-H$ and $M^{-1}$ with the resolvent ${\cal G}(z)$.
The 4 blocks result from the separation of the basis into the set $T_0$ and
the ordered set $T_1\cup T_2\cup \ldots$.
Off-diagonal matrix elements of $M$ are due to $H'$ only.
Thus $M_{11}=zI_0-H_{00}$ ($I_0$ is the unit matrix of size $t_0$ and
$H_{00}=\langle T_0|H|T_0\rangle $). $M_{22}$ is the matrix $(zI-H)$
expanded in the remaining tiers.
$M_{12}=M_{21}^t$ is a rectangular matrix of size $t_0\times
(t_1+t_2+\ldots)$.
By the tier construction, non-zero matrix elements of $M_{12}$ are
restricted to the leftmost submatrix of size $t_0\times t_1$, that
identifies with $-H'_{01}=-\langle T_0|H'| T_1\rangle$.

The aim of the calculation is to evaluate the block $(M^{-1})_{11} \equiv
G(z) \equiv G^{(0)}(z) = \langle {\bf v}_{0}|{\cal G}(z)|{\bf v}_{0}\rangle
$. The inversion formula (\ref{inv}) provides
\begin{eqnarray}
G^{(0)}(z) 
=  [zI_0-H_{00}-H'_{01} G^{(1)}(z)H'_{10}]^{-1} \,,
\end{eqnarray}
where the $t_1\times t_1$ matrix 
$G^{(1)}(z)=\langle T_1|(M_{22})^{-1}|T_1\rangle $.
To evaluate it we use equation~(\ref{inv}) again, with the blocks now resulting
from the separation of the basis into the set $T_1$ and the set $T_2\cup
T_3\cup\ldots$. Now $M_{11}=(zI_1-H_{11})$ and $M_{22}$ is the matrix
$(zI-H)$ expanded in the basis $T_2\cup T_3\cup\ldots$.  The matrix
$G^{(1)}(z)$ coincides with the block $(M^{-1})_{11}$:
\begin{eqnarray}
G^{(1)}(z) = [zI_1-H_{11}-H'_{12}G^{(2)}(z)H'_{21}]^{-1}\nonumber
\end{eqnarray}
where $G^{(2)}(z)=\langle T_2|(M_{22})^{-1}|T_2\rangle $. 
By iterating the same inversion formula (\ref{inv}) we obtain a chain of
relations of the type
\begin{equation}\label{genericinv}
G^{(k-1)}(z) = [zI_{k-1}-H_{k-1,k-1}-H'_{k-1,k}G^{(k)}(z)H'_{k,k-1}]^{-1}
\end{equation} 
%
In practice, the (in principle infinite) chain is truncated by
approximately taking $G^{(N)}(z)\approx (zI_N-H_{NN})^{-1}$.
This is the only approximation involved in this method, and it amounts to
neglecting the coupling of $T_N$ to the subsequent tier.
Starting from $G^{(N)}(z)$, one iterates (\ref{genericinv}) back to the
sought for matrix $G^{(0)}(z)$. This procedure is a matrix generalization
of the continued fraction expansion for the inversion of
tridiagonal matrices \cite{Haydock}.

%
%
%

This method provides good evaluations of the position of the quasi-stationary
states, including a rigorous treatment of all anharmonic resonances.
These effects were mostly left out in the approximate treatment of
Ref.~\cite{Stuchebrukhov93}, where off-diagonal terms $M_{12}$ in
equation~(\ref{inv}) were neglected.
The recursive calculation of the Green function (\ref{genericinv}) has
several advantages with respect to the more traditional Lanczos
method \cite{Prelovsek00,Koppel02,Wyatt98,Manini03}: (i) it provides equal
accuracy through the whole spectrum, while Lanczos method is more accurate
close to the endpoints; (ii) it splits the Hilbert space into subspaces
$T_0, ... T_N$ to treat one at a time; (iii) once the chain of matrices is
set up, each frequency requires an independent calculation, which makes
this method suitable for parallel calculations.
Its main disadvantage is the rapid growth of the tier size $t_i$, 
for systems with many degrees of freedom.
To fit the available CPU/memory limits, it is possible to cutoff the
tier growth to some maximum size $t_{\rm max}$, as described in 
Ref.~\cite{Manini03}.
In general, the recursive method may become very costly in CPU time, since
the evaluation of $G^{(0)}(E+i\varepsilon) $ requires $N$ inversions for
each sample frequency $E$, each inversion costing a time proportional to
$t_{\rm max}^3$.  In the Lanczos method, a single chain of $N_{\rm
Lanczos}\approx 10^3$ iterations, each costing of the order of the Hilbert
space size $\sim N\cdot t_{\rm max}$, generates the whole spectrum.

The \verb;c++; code for computing the tier basis and the spectrum based on the
Green-function recursive inversion formula (\ref{genericinv}) is available
in Ref.~\cite{DelMonteWeb}.

\newpage
\bibliographystyle{apsrev}

\newpage

\noindent
Table 1: Lowest saddle-point energy $E_{\rm s}$, harmonic zero-point energy
at the minimum $E_{\rm zp}(0)$, harmonic transverse zero-point energy at
the saddle point $E^\perp_{\rm zp}$, of the 4th-order PP of several
polyatomic molecules.
For diatomics, the saddle energy $E_{\rm s}$ is replaced by the height of
the maximum of the fifth-order Taylor expansion of a Morse potential
fitting the experimental harmonic frequency and dissociation energy
\cite{Bransden}.
Energies are divided by $hc$ and expressed in traditional spectroscopic
wavenumber units cm$^{-1}$.

\vskip 3mm

\begin{center}
\begin{tabular}{lrrrr}
\hline
\hline
molecular 	&	$E_{\rm s}$	&	$E_{\rm zp}({\bf 0})$	&
	$E_{\rm zp}^\perp$	&	Ref.\\
species		&	cm$^{-1}$	&	cm$^{-1}$	&
	cm$^{-1}$	&\\
\hline
N$_2$   	&	32834	&	1180	&	-	&	\cite{Bransden}\\
HCl     	&	14919	&	1495	&	-	&	\cite{Bransden}\\
H$_2$O  	&	6846	&	4717	&	4803	&	\cite{Csaszar97}\\
H$_2$O  	&	7208	&	4712	&	4488	&	\cite{Cizek93}\\
H$_2$S  	&	8529	&	3335	&	3488	&	\cite{Cizek93}\\
NO$_2$  	&	13698	&	1890	&	1936	&	\cite{Cizek93}\\
SO$_2$  	&	6263	&	1537	&	1270	&	\cite{Cizek93}\\
HOF	  	&	4624	&	3038	&	3043	&	\cite{Cizek93}\\
HOCl     	&	2821	&	2911	&	2893	&	\cite{Koput98}\\
H$_2$C$_2$O	&	834	&	6907	&	6641	&	\cite{East95}\\
C$_2$H$_4$	&	3483	&	11151	&	10815	&	\cite{Martin95}\\
CH$_3$OH	&	50	&	11398	&	11276	&	\cite{Miani00}\\
\hline
\hline
\end{tabular}
\end{center}

\newpage

\begin{figure}
\caption{\label{oneD:fig}
A typical 1-dimensional molecular potential (Morse, solid) and its
5th-order Taylor polynomial approximation (dashed) illustrating the
presence of an unbounded region separated from the physical confining
region by a barrier topping at a saddle point ${\bf q}_{\rm s}$.
}
\end{figure}

\begin{figure}
\caption{\label{water_conv:fig}
The spectrum of H$_2$O for initial excitation $|1,0,0\rangle$, computed
with $N$=3 (294 states) and $N$=15 (16811 states), broadening
$\varepsilon$=4~cm$^{-1}$.
Below the quantum energy border (dashed line) convergence is satisfactory
with relative error $\leq 1$\%.
The main features seem to converge also above $E_{\rm qb}$, but the
appearance of new structures for larger $N$ indicates non-negligible
leakage.
}
\end{figure}

\begin{figure}
\caption{\label{ketene_conv:fig}
The spectrum of ketene for initial excitation $|1,0,0,0,0,0,0,0,0\rangle$,
computed with $N$=3 (4299 states), $N$=5 (8299 states), and $N$=10 (18299
states), broadening $\varepsilon$=4~cm$^{-1}$.
No convergence is observed, even below the quantum energy border, due to
sizable leakage.
The relatively stable feature near 10000~cm$^{-1}$ corresponds to the
energy position of the initial excitation.
}
\end{figure}

\vfill
\quad
\newpage
\pagestyle{empty}

\centerline{
\epsfig{file=oneD.eps,width=135mm,clip=}
}

\vfill
\centerline{Figure 1}

\newpage
\centerline{
\epsfig{file=water_conv.eps,width=135mm,clip=}
}

\vfill
\centerline{Figure 2}

\centerline{
\epsfig{file=ketene_conv.eps,width=135mm,clip=}
}

\vfill
\centerline{Figure 3}

\end{document}